# Circumnavigating an ocean of incompressible light

## Towards quantum Hall liquids of light

Iacopo Carusotto

*INO-CNR BEC Center and Dipartimento di Fisica, Università di Trento, via Sommarive 14, I-38123 Povo, Italy*

Starting with Newton's breakthrough discovery that the same gravitational force is responsible for apples falling from trees as well as for the Moon orbiting around the Earth, a constant theme in modern physics has been that a same physical mechanism can be active in systems of hugely different size, leading to very diverse observable consequences. Rotation, for instance, is at the root of many observations in astronomical as well as condensed-matter systems, from spiral galaxies to ultra-cold atomic clouds to electron liquids in solids: on an astronomical scale, the arms of the spiral galaxy shown in the left panel of Fig.1 originate from a complex interplay of gravity, rotation and star formation in the matter forming the galaxy. On a microscopic scale, the regular arrangement of density holes in the rotating Bose-Einstein condensate shown in the right panel is a direct signature of superfluidity of the trapped atomic gas. Given the formal analogy between the Coriolis and the Lorentz forces, a most intriguing manifestation of rotation physics in a nanoscopic quantum mechanical context are the exotic incompressible phases of electron gases in strong magnetic fields with their quantized Hall resistance and the peculiar statistics of their elementary excitations.

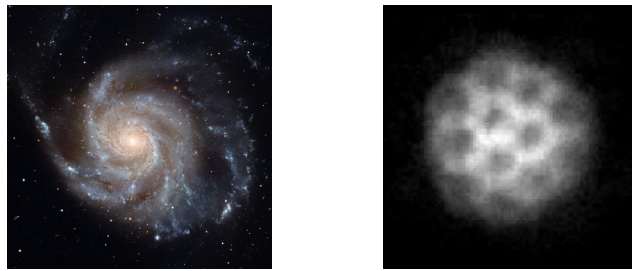

**Figure 1:**

Left panel: An example of a spiral galaxy, the Pinwheel Galaxy (also known as Messier 101 or NGC 5457). Image from http://hubblesite.org/gallery/album/entire/pr2009007h/. Credit: NASA, ESA, K. Kuntz (JHU), F. Bresolin (University of Hawaii), J. Trauger (Jet Propulsion Lab), J. Mould (NOAO), Y.-H. Chu (University of Illinois, Urbana), and STScI.

Right panel: Density profile of a rotating Bose-Einstein condensate of $^{87}$Rb atoms.
Image from F. Chevy, K. Madison, V. Bretin, J. Dalibard, cond-mat/0104218 [published in the Proceedings of the workshop *"Trapped particles and fundamental physics"* (Les Houches 2001), organized by S. Atutov, K. Kalabrese and L. Moi]

# Rotations and magnetic fields

Any undergraduate student in physics is well aware of the formal analogy between the Lorentz force felt by a charged particle moving with velocity *v* in a magnetic field B,

$F_L = q\, v \wedge B$

and the Coriolis force felt by a moving material body in a reference frame rotating at angular speed $\Omega$,

$F_C = 2m\, v \wedge \Omega,$

for instance a child bravely walking on a fast spinning merry-go-round.

What is perhaps less appreciated, is that this analogy directly extends to Hamiltonian mechanics and then to quantum mechanics. To describe the dynamics of a system in a rotating reference frame, it is in fact enough to include in the Hamiltonian the centrifugal potential

$V(r) = -1/2\, m\, \Omega^2\, r_\perp^2$

and make the minimal coupling replacement

$P \rightarrow P - A(r)$

with a vector potential

$A(r) = m\, \Omega \wedge r.$

Based on this analogy, it is then immediate to relate the Meissner effect in a superconductor to the reduced moment of inertia of a superfluid.

In both cases, the phase transition to the ordered (superfluid or superconducting) state is associated to a suppressed response of the system to an applied vector potential *A*: the *normal fraction* of the gas is in fact quantified by the linear susceptibility relating the canonical momentum density *P* to the applied vector potential *A*. In the superconductor case *A* is the real electromagnetic vector potential felt by the electrons. In the neutral superfluid case, *A* is due to mechanical rotation.

In the so-called *rotating bucket* gedanken experiment, superfluidity manifests itself as the fluid does not follow the rotation of its container and remains at rest in the inertial reference frame of the laboratory (more details in Boxes 1 and 5). In the superconductor case, one has to remind the difference between the canonical $j_{can} = qP/m$ and the physical $J_{phys} = q(P-qA/c)/m$ current densities: while superconductivity indeed fixes the canonical current density $j_{can} = 0$, the vector potential term in $j_{phys}$ is responsible for the perfect diamagnetism of superconductors in the so-called Meissner effect.

**Box 1: What is superfluidity?**

The term s*uperfluidity* is commonly used to indicate a number of different low-temperature effects that can be generically interpreted in terms of the disappearance of mechanical friction in a fluid. Superfluid behaviors were first observed simultaneously by Allen and Misener in Cambridge and Kapitsa in Moscow in 1937 as a sudden drop of mechanical viscosity when liquid 4-Helium was cooled below the so-called Lambda point at $T_\Lambda = 2.17$ K.

Following the excellent historical and conceptual introduction to superfluidity by A. J. Leggett in the cited review article, two properties that unambiguously characterize a macroscopic fluid as a superfluid are (i) the reduced moment of inertia and (ii) the metastability of supercurrents. Both of them can be physically understood in terms of the gedanken *rotating bucket experiment*.

Property (i) means that the moment of inertia of a bucket full of superfluid reduces to the one of the empty container: when the container is smoothly set into slow rotation, the superfluid remains at rest independently of the rugosity of the container walls.

Property (ii) is the time-dependent counterpart of (i): after setting the fluid into rotation by some external mean (e.g. by cooling it below the critical temperature while the container is rotating), one looks at the time it takes to the superfluid to damp out its motion once the bucket is brought at rest. While the motion of a normal fluid would be quickly damped out by friction by the container walls, a superfluid is characterized by extremely long relaxation times.

Another crucial property of superfluids involves the friction force felt by an impurity slowly moving impurity traveling through the fluid. As any cyclist knows well, friction on moving bodies can be quite substantial even in relatively rarefied fluids like air. On the other hand, impurities traveling across superfluids, e.g. ions dragged through liquid Helium by some external electric field or impurity atoms flying through atomic condensates, feel a strongly suppressed friction force as compared to the same fluid in its normal state.

# From vortex lattices to quantum Hall liquids

Of course, this picture is valid only for weak values of the vector potential intensity: when the magnetic field becomes too large, superconductivity indeed breaks down and the material goes back to its normal state. This physics is most interesting in the so-called type-II superconductors that exhibit two critical fields. Above the lower critical field $H_{c1}$, but below the upper critical field $H_{c2}$, vortices start to appear in the superconductor and organize themselves in an ordered triangular lattice, the so-called *Abrikosov vortex lattice*. As a result, perfect diamagnetism is destroyed and the magnetic field is able to penetrate the material, each vortex carrying a quantum of magnetic flux $\Phi_0 = h/(2e)$.

An analogous mechanism is active in superfluids when the fluid is set into rotation: while pioneering experiments have observed quantized vortices in liquid Helium, the most striking images of vortex lattices in superfluids were obtained in gases of alkali atoms cooled to nK temperatures using a combination of laser- and evaporative-cooling techniques. In these systems, vortices are typically created by stirring a harmonically trapped gas with an anisotropic additional potential that rotates at angular speed $\Omega$. As the stirring speed $\Omega$ approaches the trap frequency $\omega_{trap}$, more and more vortices appear in the gas and arrange in a triangular lattice analogous to the Abrikosov lattice of superconductors. An experimental picture of such a lattice is shown in the right panel of Fig.1: each vortex corresponds to a minimum in the density profile. Given the irrotationality constraint imposed by the superfluid order parameter $\psi(r)$, vorticity is concentrated at the vortex cores where $\psi(r)$ vanishes; subsequent matter wave interference experiments have confirmed that the phase of the superfluid order parameter $\psi(r)$ indeed rotates by $2\pi$ around each a vortex core. Remarkably, for large enough systems where surface effects are negligible, the vortex density $n_v = 2m\Omega / h$ is such that the fluid recovers -in average- the rigid-body rotation at $\Omega$ of a normal fluid.

Inspired by the exciting advances in our understanding of the quantum Hall effect, many authors have pushed the analogy between rotating superfluids and electron gases under strong magnetic fields further by investigating the possibility of generating more exotic strongly correlated states of atomic matter: when the rotation speed $\Omega$ gets very close to the trap frequency $\omega_{trap}$, the spatial extension of the cloud dramatically grows under the effect of the centrifugal force and the number of vortices eventually exceeds the total number of atoms $N_{at}$. In this regime, the physical meaning of the macroscopic superfluid order parameter breaks down as quantum fluctuations of the quantum matter field start to dominate over the mean-field value $\psi(r)$.

Correspondingly large quantum fluctuations are expected for the spatial motion of vortices around their equilibrium position in the Abrikosov lattice. When the spatial amplitude of this motion becomes of the order of lattice spacing, the ordered structure of the Abrikosov lattice is destroyed and the gas experiences a quantum phase transition from the superfluid state to a strongly-correlated one of the fractional quantum Hall family.

So far, severe technical difficulties have prevented this prediction from being experimentally verified: in order to obtain a strongly correlated atomic gas, the rotation speed has to be tuned with a high precision of

---

### Box 2: From dark states to synthetic gauge fields

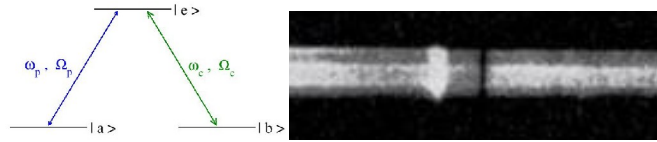

**Figure 2.** Left panel: scheme of atomic levels interacting with a bichromatic laser field. Right panel: experimental image of the black line in fluorescence spectroscopy. Image taken from the original work by G. Alzetta, A. Gozzini, L. Moi, G. Orriols, Nuovo Cimento B **36**, 5 (1976)

The concept of a *dark state* is one of the deepest and far-reaching ideas of late 20th century atomic physics. The first discovery of a dark resonance was made in Pisa in the late 1970s, when a dark line was observed in the fluorescence image of a sodium gas illuminated by a bichromatic laser field. The atomic levels involved in the optical process are arranged in a Lambda $\Lambda$ scheme and the two frequency components of the laser of frequencies $\omega_{p,c}$ and Rabi frequencies $\Omega_{p,c}$ are on resonance with the $|a\rangle \rightarrow |e\rangle$ and $|b\rangle \rightarrow |e\rangle$ transitions, respectively. In this case, there exist a quantum superposition

$$|d\rangle = \frac{1}{\sqrt{|\Omega_p|^2 + |\Omega_c|^2}} \left[ \Omega_c |a\rangle - \Omega_c |b\rangle \right]$$

of the atomic states that is completely decoupled from the laser fields as the different excitation paths to the excited $|e\rangle$ state interfere destructively. In the Pisa experiment, the position and shape of the black line was determined by the spatial dependence of the static magnetic field, which made the coherent population trapping (CPT) of atoms in the dark state $|d\rangle$ to be restricted to a given iso-B surface.

The concept of dark states was later extended to the case where the $|a\rangle$ and $|b\rangle$ states differ by the center-of-mass velocity of the atom. The subsequent velocity selectivity of the dark state underlies the VSCPT (velocity-selective coherent population trapping) mechanism to cool atoms below the recoil limit, as first demonstrated by the 1997 Nobel laureate C. Cohen-Tannoudji and his coworkers.

A different perspective on dark states was opened by Imamoglu and Harris who in 1991 demonstrated an Electromagnetically Induced Transparency (EIT) effect in a coherently dressed atomic Strontium gas: by optically dressing the $|b\rangle \rightarrow |e\rangle$ transition of the $\Lambda$ scheme with a coherent *coupling* field, resonant absorption on the other $|a\rangle \rightarrow |e\rangle$ transition could be dramatically quenched. As the linewidth of the EIT transparency line can be well below the radiative width of the bare atomic transition, extremely slow group velocities can be obtained. A dramatic slowing down of the group velocity to cycling speed in the 10 m/s regime was reported by L. V. Hau *et al.* using a Bose-condensed gas of sodium atoms and, almost simultaneously, by other authors using hot Rb gases. Following a proposal by Fleischhauer and Lukin, later experiments have extended the EIT effect to the coherent storage of light in a medium for macroscopic times.

Another fascinating application of the dark state concept was initiated by M. Olshanii and R. Dum, who realized that the spatial dependence of the laser field amplitudes $\Omega_{p,c}(r)$ leads to a spatial dependence of the dark state $|d(r)\rangle$ and, in turn, to a non-trivial Berry phase to the wavefunction of a slowly moving atom. Upon quantization of the atomic motion, the geometrical Berry phase translates into a vector potential term in the Hamiltonian, i.e. a *synthetic gauge field*.

order 1/$N_{at}$ to prevent the cloud from being destroyed under the effect of the centrifugal potential. On the other hand, the cylindrical symmetry of the trap has to be precise enough for the large amount of angular momentum stored in the fast rotating gas not to be quickly damped away by experimental imperfections. Thanks to the recent advances in the experimental techniques, a promising strategy is to work with small samples containing a very small number of atoms for which both constraints are less severe. Experimental investigations in this direction are presently very active.

## Synthetic gauge fields

In the last years, several groups have started exploring a completely different approach to orbital magnetism in neutral atomic gas using the so-called *synthetic gauge field* concept. As it is historically reviewed in Box 2, the basic idea of a synthetic gauge field is a consequence of the *dressed atom* concept: as the characteristic timescale of the internal and external degrees of freedom of the atom is very different, a Born-Oppenheimer approximation of the atomic center-of-mass motion can be developed, where the coupling to the internal degrees of freedom results is summarized by a vector potential term $A(r,t)$ to be included in the Schrödinger equation. Historically, this idea has been long known in molecular physics, but was later independently rediscovered by Dum and Olshanii in 1996 and applied to atomic physics. In this latter context, a wide flexibility on the spatial and temporal shape of the vector potential $A(r,t)$ can be obtained by simply playing with the amplitudes and frequencies of the dressing lasers or even by imposing additional static fields to the atoms.

The use of synthetic gauge fields instead of a global rotation of the atomic cloud offers several interesting practical advantages in the quest for strongly correlated states of atomic matter: the fact that the atoms are not physically moving in the laboratory frame significantly reduces the detrimental effect of trap asymmetries. Furthermore, synthetic gauge fields can be made arbitrarily large without making the cloud to fly apart under the effect of the centrifugal potential that is normally associated to rotations.

First implementations of this idea was demonstrated in the group of I. Spielman at JQI in Maryland, who showed how by suitably tailoring the synthetic gauge field in both space and time one can generate spatially homogeneous synthetic magnetic and electric fields. While the effect of the synthetic electric field is to accelerate the atoms according to a synthetic version of the Faraday law, the effect of the synthetic magnetic field onto a Bose-condensed cloud is to generate quantized vortices in the cloud in a very similar way to what was observed upon rotating the cloud. Also in this case, however, the number of vortices generated in the cloud remains orders of magnitude lower than the total number of atoms and strong experimental efforts are still needed before the quantum Hall regime can be penetrated.

**Box 3: Massive photons in planar microcavities**

Since our early studies, we are taught of the photon being massless particle. This is of course true in vacuo, but ... what happens under spatial confinement, e.g. between a pair of plane-parallel mirrors?

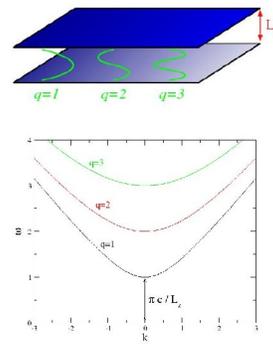

**Figure 3:** sketch of a planar cavity and of the dispersion of its e.m. modes

In the idealized case of a pair of perfect metallic mirrors separated by a distance $L_{zz}$, the electromagnetic modes can be found by imposing the boundary condition that the electric vanishes at the surface of the metal. This results in the wavevector component orthogonal to the mirror planes being quantized as $k_z = q\,\pi/L_{zz}$, q being a positive integer q>0. On the other hand, the in-plane components remain free to take any values $k_x, k_y$.

Inserting this ansatz into Maxwell's equations, one immediately finds that the electromagnetic modes organize in branches, each of them with a relativistic dispersion of the form

$$\omega = c\sqrt{k_x^2 + k_y^2 + \frac{q^2 \pi^2}{L_z^2}}$$

.For the most commonly used lowest q=1 branch, the rest-mass of the cavity photon is given by

$$m^* = \frac{\hbar \pi}{c L_z}$$

modulo a factor 2, the cavity length $L_{zz}$ then plays the role of the Compton wavelength $\lambda_c = h/m^*c \sim 2L_{zz}$ for the cavity photon.

## Superfluids of light

While the atomic physics community is intensely working to generate strongly correlated quantum Hall liquids of atoms, another example of quantum fluid is making its first steps in the world of many-body physics.

Historically, we are used to consider matter and light as two very distinct entities: we are able to feel the presence of matter with our hands; in its solid version, matter can be remarkably impenetrable and even in its fluid versions, it can sustain the weight of heavy objects swimming and flying through it. In contrast, our common perception of light is like an impalpable substance that flies from objects to our eyes and makes them visible to us. We are of course well aware of special conditions where the pressure of light has observable consequences, such as the long tail of comets, or it is even exploited in the lab to laser-cool atoms down to nanoK temperatures. Still, our intuitive picture is that the photons forming a light beam are emitted by the source, travel along a straight line across space until they are either absorbed or scattered by matter. The dual wave-particle nature of the quantum mechanical photon only adds interference features to this simple propagation picture, but does not question the overall intuitive image.

This is no longer the case of recent experiments demonstrating situations where the many photons forming a light field propagating in a nonlinear optical medium and/or confined in a planar microcavity behave in a collective way as a fluid, the so called *fluid of light*. As it is discussed in Box 4, the Kerr optical nonlinearity of the optical medium induces sizable interactions between individual photons. Provided the resulting photon-photon collision are frequent enough, the photon gas may even reach an independent thermal equilibrium state at a temperature different from the one of the surrounding environment.

A striking example of collective behavior in a photon gas are the recent observations of Bose-Einstein condensation effects in optical systems. Several systems have been used to this purpose, from vertical cavity surface emitting lasers, to semiconductor microcavities in the strong coupling regime, to macroscopic optical cavities filled with dye molecules, to simple nonlinear optical crystals. Common features shared most of them are the finite rest mass that is attributed to the photon by the spatial confinement of the electromagnetic field in cavity (see Box 3) and the importance of stimulated scattering processes that tend to accumulate particles into a single single-particle state.

While the typical smoking gun of BEC in dilute atomic gases is a narrow peak at $k=0$ in the momentum distribution, the signatures of BEC in optical systems closely remind the spontaneous coherence usually observed in laser emission: a tight angular collimation of the emission, long-range spatial and temporal coherence, and suppressed intensity and phase noise. From this perspective, the actual distinction between lasing and BEC is rather elusive and many authors -including the writer- are very much inclined to view photon BEC and lasing as different examples of the same second-order phase transition spontaneously breaking the U(1) symmetry of phase rotations.

Of course, there remains a crucial difference with respect to standard textbook quantum statistical mechanics: unless the thermalization time of photons is really much faster than the photon lifetime in the device, the state of the photon gas is determined by a dynamical balance of pumping and losses, rather than by the thermal equilibrium condition. To stress this important issue, one should then speak of lasing as an example of *non-equilibrium Bose-Einstein condensation*. While the critical properties of this phase transition are still under active study, several important features of the dynamical behavior of the condensate have been experimentally investigated in the last years.

In this perspective, planar microcavity samples in the strong light-matter coupling regime are among the most powerful workhorses, as mixing the cavity photon with a narrow excitonic transition into new normal excitations -the so-called *exciton-polaritons*- dramatically enhances the effective optical nonlinearity of the system while maintaining a low polariton mass and an efficient radiative coupling to the external world. The first comprehensive studies of BEC phenomena in the optical context were indeed performed in this kind of samples.

**Box 4: Photon-photon interactions**

While the linear form of Maxwell's equations of classical electrodynamics predicts no interactions between photons propagating in vacuum, a remarkable consequence of quantum electrodynamics is that vacuum displays a non-vanishing optical nonlinearity due to virtual electron-positron pairs according to the Feynman diagram in figure.

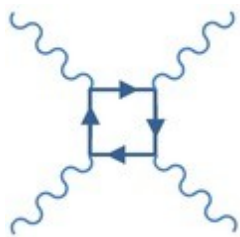

This result was first derived by Heisenberg and Euler and leads to a finite photon-photon scattering cross section

$$\sigma \sim \alpha^4 \frac{\hbar^2}{m^2 c^2} \left( \frac{\hbar \omega}{mc^2} \right)^6$$

Given the small value of the electron Compton wavelength $\hbar/mc = 0.4$ pm and the large electron mass m=0.5MeV, this cross-section turns out to be ridiculously small for photons in the eV range of optical interest, and the process has never been observed.

The situation is of course very different in optical media, where optical transitions are available in the same eV energy range of the propagating photons: instead of creating virtual electron-positron pairs in vacuo, one can in fact create electron-hole pairs in a semiconductor or excite an internal transition of an atom. In both cases, the $mc^2$ denominator is dramatically reduced and the photon-photon cross-section becomes experimentally observable.

Such nonlinear optical phenomena were first observed using intense laser beams where a huge number of photons can actually participate to the process and the rate of the process may be dramatically enhanced by bosonic stimulation. Among other, this is the case of second harmonic generation, optical bistability, optical parametric oscillation.

The new frontier is now to observe scattering between optical wavepackets containing single photons: this process is for instance at the heart of the photon blockade mechanism discussed in the main text. If properly mastered in optical networks may lead to breakthrough application in all-optical quantum computing.

Very soon after the observation of BEC, researchers started wondering whether superfluidity effects could be observed in Bose-condensed quantum gases of photons (or equivalently polaritons) in planar semiconductor microcavities. A proposal in this direction was put forward by the writer in collaboration with Cristiano Ciuti in 2004 and was later implemented by the LKB-Paris 6 group in Paris led by Alberto Bramati and Elizabeth Giacobino. Fig.4 shows some experimental images from this experiment. A coherent gas of polaritons is flowing along the plane of a microcavity and hits a localized defect naturally

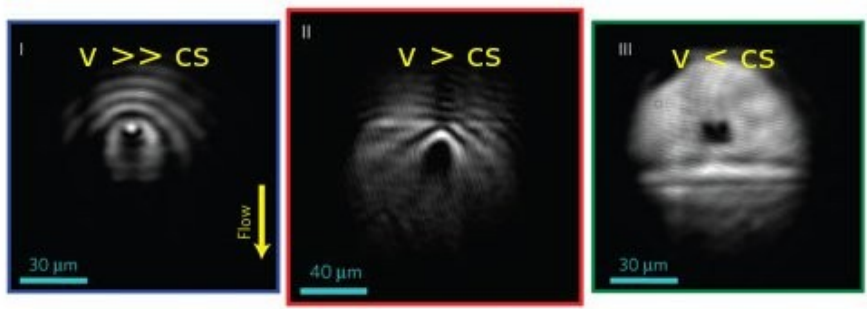

**Figure 4**: Experimental images of a quantum fluid of light hitting a localized structural defect at different speeds. decreasing from left to right. Images from A. Amo et al., Nat. Phys. **5**, 805 (2009)

present in the microcavity sample.

The different panels of the figure refer to different ratios of the flow speed $v$ (directed in the downwards direction in the figure) over the sound speed $c_s$ in the photon gas. The left panel is for a fast flow $v/c_s \gg 1$: photons are elastically scattered by the defect in all directions; their interference with the in-coming photons creates the parabolic-shaped fringes that extend mostly in the upstream direction from the defect. The central panel is for a moderately super-sonic flow $v/c_s > 1$: in addition to the parabolic fringes upstream of the defect, one clearly sees a triangular feature located right downstream of the defect. This is the superfluid analog of the Mach cone that is created in air by a supersonically flying jet.

Finally, the right panel is for a sub-sonic flow $v/c_s < 1$: in this case, the defect is no longer able to produce any propagating disturbance in the fluid and, the density modulation reduces to a localized dip in the neighborhood of the defect. This is the smoking gun of superfluid behavior in the polariton gas. If at some point of the next future it becomes possible to probe superfluidity using mobile impurities instead of structural defects, the superfluid regime will be signaled by a sudden drop of the radiation pressure force exerted by the flowing polaritons onto the defect.

Almost in the same years, a number of other groups have started exploring other features of superfluidity in polariton fluids. A full account of these exciting advances can be found in the cited review article by the writer and C. Ciuti. Among the most intriguing results, we may mention here the observation of oblique dark solitons and vortex nucleation at the surface of

**Box 5: Metastable supercurrents of light in aircrafts**

While a direct evidence of metastable supercurrents of light along the plane of a semiconductor microcavity was only recently reported by the UAM group [D. Sanvitto et al., Nature Physics 6, 527 (2010)], a closely related effect has been exploited in gyrolasers for aerospace applications.

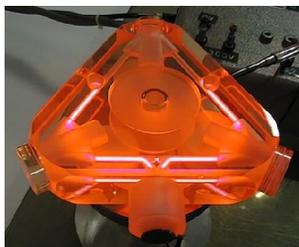

**Figure 5:** photo of a gyrolaser. © Thales, courtesy S. Schwartz.

To understand this analogy in depth, one has to remind how the usual quantum mechanical description of metastable superflow in a toroidal container involves the conservation of the so-called winding number, i.e. the number of $2\pi$ loops performed by the phase of the superfluid order parameter while going around the torus. Unless some extra spin degrees of freedom are present, any change in the winding number requires creating a hole in the superfluid density, which involves a relatively large energy cost.

In ring cavities, the role of the winding number is played by the discreteness of optical modes, labeled by the number of optical wavelengths contained in a round trip. When such a ring cavity is filled with an amplifying medium, we know that laser operation occurs on a given cavity mode. In order for the winding number to change in a so-called mode jump, a strong perturbation of the electric field is needed.

Actual operation of a gyrolaser is based on the different Sagnac shift experienced by different modes of a ring laser when the aircraft on which the device is mounted is performing a turn. Provided one manages to force the device to simultaneously lase on two counter-propagating modes, rotation of the aircraft is then directly read in the beat note between the two emissions.

Superfluidity is then essential to this gyrolaser operation, as it guarantees that the laser device does not perform mode jumps and the only effect of rotation is a frequency shift of the different lasing modes.

a spatially large and impenetrable impurity in a polariton flow and the characterization of long-lived polariton supercurrents in toroidal geometries (see Box 5 for more details). While these features closely resemble what one already knew from traditional superfluid hydrodynamics of liquid Helium and dilute atomic condensates, many new features have been theoretically anticipated stemming from the very non-equilibrium -- driven-dissipative -- nature of the photon fluid. As a most striking consequence of pumping and losses, we may mention that in many circumstances propagation of sound in a polariton fluid is not allowed but is replaced by an overdamped Goldstone mode with a diffusive dispersion.

## Strongly-correlated photon gases

Even though the viewpoint is markedly different, most of the physics of photon condensation and photon superfluidity discussed in the previous section might have been discussed decades ago from a purely nonlinear optical standpoint: the Gross-Pitaevskii equation for the photon condensate is in fact a rewriting from a different perspective of Maxwell equations for coherent light including a nonlinear polarization term within the paraxial approximation. The discrete nature of the individual photon forming the gas only becomes instead essential when one enters the so-called *strongly interacting photon* regime.

This regime is easy to understand in the simplest case of a single-mode cavity containing a strongly nonlinear medium. As usual, the $\chi^{(3)}$ nonlinear susceptibity of a Kerr medium makes the effective refractive index of the cavity to depend on the light intensity. In classical nonlinear optics, the resulting frequency shift of the cavity resonance frequency can be exploited to observe a number of nonlinear optical effects, from optical limiting to optical bistability.

Marked departures from classical optics are instead observable when the frequency shift given by a single quantum of light intensity exceeds the linewidth of the cavity resonance. In this case, while a first resonant photon can freely penetrate the cavity, the entrance of a second photon is forbidden by the nonlinear shift of the cavity frequency due to the presence of the first photon. This phenomenon was originally proposed by Atac Imamoglu and coworkers. Given the close analogy with Coulomb blockade of mesoscopic conductors, it generally goes under the name of *photon blockade*. Its most direct consequence is a strong photon antibunching displayed by the light beam transmitted through the cavity.

When many such cavities are coupled to each other, a kind of Bose-Hubbard model is obtained, where the fundamental particle is the photon and each cavity plays the role of a lattice site: photon blockade then provides an impenetrability condition preventing more than one photon from sitting on the same site.

In the limiting case where the photon lifetime is much longer than all other time scales of the problem, a superfluid to Mott-insulator transition has been predicted to occur in a photon gas confined in a lattice of non-linear coupled single-mode cavities: while the photons of a superfluid are delocalized over the whole system and enjoy long-range spatial and temporal coherence, particles in a Mott-insulator state are spatially localized; spatial coherence is replaced by dramatic suppression of number fluctuations and each site ends up containing a fixed integer number of particles. Unfortunately, the unavoidable finite photon lifetime makes this transition hard to observe in practice: the loss of just a few photons is in fact enough to destabilize a Mott insulator and turn the system back to the more standard superfluid state.

But Mott-insulators are not the only interesting many-body states of impenetrable photons. A remarkable state of matter was predicted by Girardeau in 1960 for impenetrable bosons in a one dimensional geometry: in this case, bosons turn out to be effectively "fermioned", that is acquire many of the properties of a non-interacting Fermi gas such as a Fermi pressure in the equation of state and the Pauli hole in the density-density correlation function. Most of these results are easily theoretically described thanks to a mathematical mapping of the bosonic many-body wavefunction onto the one of a related fermionic system. Theoretical work for the photonic case has shown how the properties of such a Tonks-Girardeau gas are robust against losses and how the coupling with the radiative environment can be rather exploited to study many-body states in resonant transmission experiments. As compared to the experiments with ultra-cold atomic gases where Tonks-Girardeau physics is often demonstrated in an indirect way, the photonic experiment would provide direct access to all quantum coherence properties of the gas just by looking at the coherence properties of the emitted light.

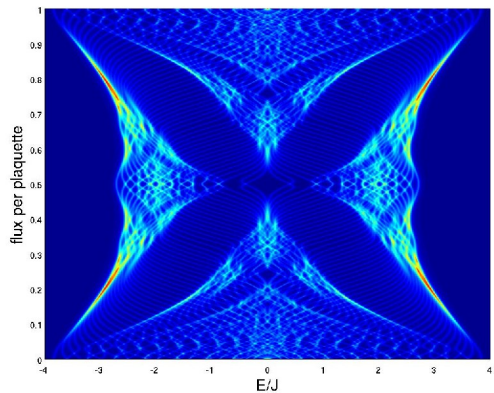

**Figure 6:** Hofstadter butterfly shape in the transmission diagram of an array of optical cavities in the presence of a synthetic gauge field for light.

While photon blockade in single mode cavities has been observed by several groups using either single atoms or single quantum emitters in solid-state devices, preliminary evidence of strong photon-photon interactions in a spatially extended, many-mode context has been recently reported by the group led by Mikhail Lukin and Vladan Vuletic at Harvard. Once again, the EIT phenomenon turns out to be at the heart of an important discovery: the observed optical nonlinearity is based on a three-level scheme with a Rydberg atom state as the final state. As soon as two photons sit closer than a blockade radius of the order of 10 microns, the long-range interaction between the huge electric dipole of atoms in the Rydberg state is able to spoil EIT and restore absorption. As it is signaled by the marked anti-bunching in the emerging light beam, the atomic gas behaves as an optical switch whose absorption level is controlled by the presence of even a single photon.

## Synthetic gauge fields for light

The physics of quantum fluids of light becomes even more exciting when a global rotation is imposed to the system and/or a synthetic gauge field for photons is generated. Rotating photon gases are routinely created by illuminating a cylindrically symmetric many-mode cavity with a pump beam with a Laguerre-Gauss profile with finite orbital angular momentum, so to create e.g. arrays of optical vortices. On the other hand, synthetic gauge fields for photons can be generated by suitably tailoring the optical set-up so to induce geometrical phases in light propagation.

Historically, the first example of such geometrical phase dates back to the work of Pancharatnam who realized that propagation of circularly polarized light through a sequence of $\lambda/2$ slabs with different orientation induces a phase shifts that changes sign when the direction of light propagation is reversed. This very simple classical optics phenomenon has an interesting geometrical interpretation on the Bloch sphere of polarizations as the Berry phase that acquired by a quantum mechanical particle when the orientation of its spin is slowly rotated along a non-trivial path.

More recently, a number of set-ups have been proposed to exploit such geometrical phases to realize arrays of optical cavities with non-trivial tunneling phases, which corresponds to imposing a synthetic gauge field to the photon motion. Among the most remarkable proposals in this direction, we may mention arrays of silicon whispering-gallery microcavities connected by an asymmetric pair of waveguides, stripline microwave cavities coupled via rings of Josephson junctions, and optical cavities with a suitable time-modulation of their oscillation frequency. In all cases, the basic element needed to generate the synthetic gauge field is some violation of time-reversal symmetry. Analogously to carriers propagating along the edge of incompressible quantum Hall liquids of electrons, a direct observable consequence of the synthetic gauge field is the appearance of chiral edge states unidirectionally propagating along the surface of the system, A first experimental observation of such topological effects in microwave propagation in a magneto-optical photonic crystal is discussed in Box 6. Several other experiments investigating synthetic gauge fields for light are presently under way both in the microwave and in the optical domains.

### Box 6: An (integer) quantum Hall effect with light

The first example of quantum Hall effect for light was recently demonstrated by Marin Soljacic's group at MIT. Following a pioneering prediction by Haldane and Raghu, they constructed a two-dimensional photonic crystal structure embedding ferrite rods and investigated the propagation of microwaves through it.

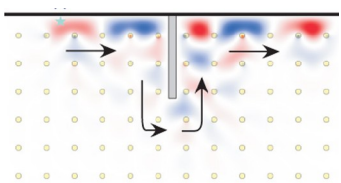

**Figure 7:** Spatial profile of an e.m. field unidirectionally propagating along the edge of a magneto-optical photonic crystal and wrapping around a defect without being backscattered. Figure from Z. Wang et al., Nature **461**, 772 (2009)

Photonic crystals are under active study since the '80s as they allow for a variety of novel light propagation effects: given their spatially periodic structure, photonic states in their bulk are organized in allowed bands and forbidden gaps in very much the same way as it occurs to electrons in crystalline solids. The main novelty of the device used in the present experiment is that the gyrotropic permeability of the ferrite rods breaks time-reversal symmetry and makes the photonic bands to acquire a non-trivial topology.

As it is typical in the field of *topological materials* (insulators or even superconductors), the term "topology" indicates here a quite sophisticated geometrical property of an allowed energy band, whose precise definition involves the Chern number of a connection -the Berry connection- defined on a manifold with the structure of the first Brillouin zone of the lattice. As a direct consequence of the non-trivial topology of the bands, photonic edge modes must appear at frequencies within the forbidden gap of the bulk crystal and spatially localized on the geometrical surface of the system.

The most exciting property of these edge states is their well-defined chirality, that is their ability to propagate along the edge in one direction only. As a result, all back-scattering processes on disorder are efficiently suppressed: as it is illustrated in the figure, whenever a defect is present along its path, light just wraps around it and then keeps propagating as if nothing had happened. Research in view of exploiting this physics as novel components of photonic networks is presently very active.

Fascinating consequences of the synthetic gauge field on the transmission spectrum of an array of cavities are illustrated in Figure 6: plotted as a function of the incident laser frequency and of the synthetic gauge field strength, the regions of high transmittivity organize in the fractal geometrical structure called the *Hofstadter butterfly* after the name of D. Hofstadter who first discussed it in the context of electron states in crystals under huge - almost megaT - magnetic fields. For rational values *p/q* of the synthetic gauge field flux in units of a flux quantum per lattice plaquette, high transmittivity is found in *q* frequency windows. Each of them corresponds to a photonic band. As a result of the non-trivial topology of these bands, the gaps between the bands are crossed by edge states with well-defined chirality, analogous to the ones observed in the MIT experiment discussed in Box 6.

# Why the title?

We wish to conclude our journey across the exciting perspectives of quantum fluids of light with a few words explaining the title of this article... what is an ocean of incompressible light? what is the interest in circumnavigating it?

The answer again lies in the analogy with the quantum Hall physics of two-dimensional electron gases under strong magnetic fields and, more in general, with topological states of matter. The exciting challenge is to investigate what new physical effects can be observed by combining strong photon-photon interactions with strong synthetic gauge field. More specifically, could one envisage to generate new quantum phases of the photon fluid, perhaps with incompressibility properties like fractional quantum Hall liquids? How does the non-equilibrium nature of the photon fluid affect the quantum Hall physics? is it possible to take advantage of the peculiar anyonic statistics of edge excitations in quantum information processing devices that are topologically protected from decoherence? Nobody knows yet... and our curiosity is very strong! Of course, many obstacles are still present along this ambitious route, and are most likely to keep on challenging physicists for decades. But as usual, there is no other way to find it but to fill the hold with all tomes of the Landau-Lifshitz series (and perhaps some food and fresh water) and quickly sail out in open sea!!

## The author

IC completed his PhD in 2000 at Scuola Normale Superiore in Pisa under the supervision of prof. G. C. La Rocca. After a post-doc in Paris at LKB in the group of Y. Castin, since 2003 he works in Trento as a Researcher at the BEC Center of INO-CNR. During the years, he has been Maitre de Conferences at College de France in the group of C. Cohen-Tannoudji and Visiting Professor and ETH in the group of A. Imamoglu. His research interests range from nonlinear and quantum optics of the so-called quantum fluid of light, to ultracold atomic gases and condensed-matter models of quantum field theories on curved space-times. The research that is presented in this article would not have been possible without the continuous support of many colleagues, in particular Cristiano Ciuti, Atac Imamoglu, Rifat Onur Umucalilar and Michiel Wouters. The authors is indebted to G. La Rocca, M. Artoni, F. Bassani and M. Inguscio for introducing him to the physics of dark states and Electromagnetically Induced Transparency. The interested reader can find a complete description of the on-going research activities at the BEC Center on the author's homepage www.science.unitn.it/~carusott and on the group's webpage bec.science.unitn.it.

## Further reading